\documentclass[aps,print,showpacs,showkeys,10pt,twocolumn]{revtex4}%
\usepackage{amsfonts}
\usepackage{amsmath}
\usepackage{amssymb}
\usepackage{graphicx}%
\providecommand{\U}[1]{\protect \rule{.1in}{.1in}}
\begin{document}
\title{Nonautonomous Bright and Dark Solitons of Bose-Einstein Condensates with
Feshbach-Managed Time-Dependent Scattering Length}
\author{Qiu-Yan Li$^{1,2}$, Zai-Dong Li$^{1,2}$, Lu Li$^{3}$,}
\email{llz@sxu.edu.cn}
\author{Guang-Sheng Fu$^{2}$}
\affiliation{$^{1}$Department of Applied Physics, Hebei University of Technology, Tianjin
300401, China}
\affiliation{$^{2}$School of Information Engineer, Hebei University of Technology, Tianjin,
300401, China}
\affiliation{$^{3}$College of Physics and Electronics Engineering, Shanxi University,
Taiyuan, 030006, China}

\pacs{03.75.Lm, 05.30.Jp, 67.40.Fd}
\keywords{Nonautonomous soliton solution; interaction; Bose-Einstein condensation}
\begin{abstract}
We present a family of nonautonomous bright and dark soliton solutions of
Bose-Einstein condensates with the time-dependent scattering length in an
expulsive parabolic potential. These solutions show that the amplitude, width,
and velocity of soliton can be manipulated by adjusting the atomic scattering
length via Feshbach resonance. For the cases of both attractive and repulsive
interaction, the total particles number is a conservation quantity, but the
peak (dip) density can be controlled by the Feshbach resonance parameter.
Especially, we investigate the modulation instability process in uniform
Bose-Einstein condensates with attractive interaction and nonvanishing
background, and clarify that the procedure of pattern formation is in fact the
superposition of the perturbed dark and bright solitary wave. At last, we give
the analytical expressions of nonautonomous dark one- and two-soliton
solutions for repulsive interaction, and investigate their properties analytically.

\end{abstract}
\maketitle

\section{Introduction}

The classical soliton concept was introduced firstly by Zabusky and Kruskal
\cite{Zabusky} for autonomous nonlinear and dispersive dynamic systems where
the time variable has only played the role of an independent variable and has
not appeared explicitly in the coefficients of the nonlinear evolution
equation. These autonomous solitons do not disperse and completely preserve
their localized form and speeds during propagation which has motived a great
attention in optical fibers and condensate physics. As a theoretical model,
the nonlinear Schr\"{o}dinger equation has been adopted extensively to govern
the dynamics of autonomous bright and dark solitons in optical fibers
\cite{Hasa,Ablowitz} and Bose-Einstein condensates (BECs) \cite{Dalfovo}.

In BECs bright and dark solitons have been paid more particular interest
experimentally and theoretically. The bright soliton
\cite{Strecker,Carr,Sal,Khawaja,Kev,lu06,Wu} is expected by the balance
between dispersion and attractive mean-field energy. However, collapse of
bright soliton \cite{Cornish} may occur owing the attractive interaction of
bosons. To avoid this collapse, one should restrict the BECs dynamics into the
quasi-one-dimensional regime, i.e., the energy of two body interaction is much
less than the kinetic energy in the transverse direction. The dark soliton
\cite{Dum,Burger,Jackson,Busch,Bronski} can be formed in the case of repulsive
interaction of bosons. It denotes the macroscopic excitation characterized by
a local density minimum, and a phase gradient of the wave function at the
position of the minimum. Comparing with the attractive case, large condensates
can be realized for dark soliton owing the repulsive interaction of bosons.

When the physical system is subjected to various external time-dependent
forces, the nonautonomous nonlinear evolution models typically arise and the
term of nonautonomous solitons \cite{Serkin3} was introduced firstly. In fact,
different aspects of dynamics in nonautonomous models \cite{Chen,Konotop,Li2}
and the controllable soliton solutions \cite{Serkin3,Serkin,Serkin2,LiLu} in
optical fibers have been investigated theoretically. For BECs the nonlinearity
resulting from the interatomic interaction is denoted by the effective
scattering length which can be tuned experimentally by utilizing the Feshbach
resonance \cite{Roberts}, even including its sign. A sinusoidal variation of
the scattering length has also been used to form patterns such as Faraday
waves \cite{Staliunas,Engels}. The controlling soliton trains \cite{Abdullaev}
was also created starting from periodic waves. Therefore, the BECs solitons
forming by magnetically tuning the interatomic interaction near a Feshbach
resonance are a typical example of such nonautonomous solitons in external
potentials and offer a good opportunity for the nonlinear excitations
exploration \cite{Pelinovsky}. The properties of such nonautonomous solitons
in BECs are not well explored and it is our purpose in the present paper.

At the mean-field level, the evolution of the macroscopic wave function of
BECs can be described by the Gross-Pitaevskii equation
\cite{Kevrekidis2,Brazhnyi},%
\begin{equation}
i\hbar \frac{\partial \Psi \left(  \mathbf{r,}t\right)  }{\partial t}%
=[-\frac{\hbar^{2}\triangledown^{2}}{2m}+V_{\text{ext}}\left(  \mathbf{r}%
\right)  +g\left \vert \Psi \left(  \mathbf{r,}t\right)  \right \vert ^{2}%
]\Psi \left(  \mathbf{r,}t\right)  , \label{gp}%
\end{equation}
where $\Psi \left(  \mathbf{r,}t\right)  $ is normalized to the number of
condensed atoms, i.e., $N=\int \left \vert \Psi \right \vert ^{2}d^{3}r$, $m$ is
the atomic mass, $V_{\text{ext}}$ is a harmonic trap given by $V_{\text{ext}%
}\left(  \mathbf{r}\right)  =m/2[\omega_{0}^{2}x^{2}+\omega_{\bot}^{2}\left(
y^{2}+z^{2}\right)  ]$ with $\omega_{0}$ and $\omega_{\bot}$ being the axial
and transverse harmonic oscillator frequencies, and the effective interatomic
interaction reads $g=4\pi \hbar^{2}a_{s}/m$ with $a_{s}$ being the $s$-wave
scattering length ($a_{s}<0$ for attractive interaction; while $a_{s}>0$ for
repulsive interaction). The linear oscillator lengths in the transverse and
cigar-axis directions is defined by $l_{\bot}=(\hbar/m\omega_{\bot})^{1/2}$
and $l_{0}=(\hbar/m\left \vert \omega_{0}\right \vert )^{1/2}$, respectively.
For a cigar-shaped condensate at a relatively low density, when the energy of
two body interactions is much less than the kinetic energy in the transverse
direction, i.e. when $\epsilon^{2}=\left(  l_{\bot}/\xi \right)  ^{2}\sim
N\left \vert a_{s}\right \vert /l_{0}\ll1$ \cite{Brazhnyi}, where $\xi=\left(
8\pi n\left \vert a_{s}\right \vert \right)  ^{-1/2}$ is the healing length and
$n\propto N/\left(  l_{\bot}^{2}l_{0}\right)  $ is a mean particle density,
then the system becomes effectively quasi-one-dimensional regime
\cite{Strecker,liang} with the explusive potential%
\begin{align}
i\frac{\partial}{\partial t}\psi \left(  x,t\right)   &  =-\frac{\partial^{2}%
}{\partial x^{2}}\psi(x,t)-\frac{1}{4}\lambda^{2}x^{2}\psi(x,t)\nonumber \\
&  +2N\frac{a_{s}}{l_{\bot}}\left \vert \psi(x,t)\right \vert ^{2}\psi(x,t),
\label{nls1}%
\end{align}
where $\lambda \equiv2\left \vert \omega_{0}\right \vert /\omega_{\bot}\ll1$, and
the time $t$ and coordinate $x$ has been measured in units $2/\omega_{\bot}$
and $l_{\bot}$, respectively. In real experiment for BECs \cite{Strecker} the
scattering length $a_{s}$ is tuned with a Feshbach resonance from repulsive to
attractive. For soliton created with the particle number $N\sim10^{3}$,
$\omega_{\bot}=$ $2\pi \times710$Hz, $\omega_{0}=2\pi i\times20$Hz, the
parameter $\lambda=0.056$ is small. With the above conditions, the units
$l_{\bot}\ $is about $1.44\mu m$ and the units time for a BECs trapped with
the transversal size order $l_{\bot}$ corresponds to $4.5\times10^{-4}$s. The
lifetime of a BECs is of the order of $1$s, which is about 220 in our
dimensionless units. If one chose the scattering length is increased in the
form of $a_{s}\left(  t\right)  =a_{0}\exp(\lambda t)$ with the initial
scattering length $a_{0}=-0.02$ nm. After 40 dimensionless units of time, the
value of the atomic scattering length turns to $a_{s}=-0.19$ nm corresponding
to $\epsilon^{2}\sim0.02$ which provides the safe range parameters.

In terms of the transformation $\psi \left(  x,t\right)  =q\left(  X,T\right)
\exp \left(  \lambda t/2-i\lambda x^{2}/4\right)  $ with $X=x\exp \left(
\lambda t\right)  $ and $T=2\int_{0}^{t}\exp \left(  2\lambda \tau \right)
d\tau$, Eq. (\ref{nls1}) reduces to the standard form%
\begin{equation}
i\frac{\partial q}{\partial T}+\frac{1}{2}\frac{\partial^{2}q}{\partial X^{2}%
}-\left(  Na_{0}/l_{\bot}\right)  \left \vert q\right \vert ^{2}q=0.
\label{nls2}%
\end{equation}
This result show that in BECs the nonautonomous solitons formed by
magnetically tuning the interatomic interaction via Feshbach resonance can be
obtained from the autonomous solitons. In this paper, we will explore the
generalized nonautonomous bright and dark solitons of Eq. (\ref{nls1}). As an
example, the nonautonomous bright soliton solutions are obtained on the vacuum
state background and nonzero background, respectively. With these exact
solutions the corresponding dynamic properties are discussed in detail. At
last, we investigate the dynamic behavior of the nonautonomous dark soliton solutions.

\section{Nonautonomous bright soliton solutions}

In this section, we consider the case of attractive interaction between atoms,
i.e., $a_{0}<0$. In this case, Eq. (\ref{nls2}) is a integrable model, and its
soliton solutions can be constructed by several technique, such as inverse
scattering method \cite{Ablowitz}, Darboux transformation
\cite{Xu,Lilu,Lishuqing} and Hirota method \cite{Hirota}. With the expulsive
parabolic potential and the Feshbach-managed time-dependent scattering length,
we will present the exact nonautonomous bright soliton solutions analytically
on the vacuum state background and nonzero background for Eq. (\ref{nls1}).

\subsection{Nonautonomous bright soliton solution and soliton interaction on
the vacuum state background}

Firstly, we can present the nonautonomous one-soliton solution for Eq.
(\ref{nls1}) as follows
\begin{equation}
\psi=A_{s}\operatorname{sech}\theta_{s}\exp \left(  i\varphi_{s}+\lambda
t/2\right)  ,\label{bso1}%
\end{equation}
where $\theta_{s}$ and $\varphi_{s}$ are given by%
\begin{align}
\theta_{s} &  =\mu A_{s}\left(  xe^{\lambda t}-2k_{s}\int_{0}^{t}%
e^{2\lambda \tau}d\tau \right)  -\theta_{0},\nonumber \\
\varphi_{s} &  =k_{s}xe^{\lambda t}-\frac{\lambda x^{2}}{4}+\left(  \mu
^{2}A_{s}^{2}-k_{s}^{2}\right)  \int_{0}^{t}e^{2\lambda \tau}d\tau-\varphi
_{0},\label{para1a}%
\end{align}
here $\mu=\sqrt{N\left \vert a_{0}\right \vert /l_{\bot}}$. The solution $\psi$
in Eq. (\ref{bso1}) describes a bright soliton of BECs with time-dependent
atomic scattering length in an expulsive parabolic potential, with the initial
maximum amplitude $A_{s}$, the initial wave number\textbf{\ }$k_{s}$, the
initial location $\theta_{0}/(\mu A_{s})$, and the initial phase $\varphi
_{0}/k_{s}$. When $\lambda=0$, Eq. (\ref{bso1}) can reduce to the solution for
the standard nonlinear Schr\"{o}dinger equation. From Eq. (\ref{para1a}) we
see that the soliton in Eq. (\ref{bso1}) can undergo compressing effect with
the increasing of the scattering length tuned by the Feshbach resonance. This
character does not exist for the situation of the uniform nonlinear
Schr\"{o}dinger equation \cite{Ablowitz}. This nonautonomous soliton has an
increase in the peak value, while the particles number is a conservation
quantity due to $\int_{-\infty}^{+\infty}\left \vert \psi_{1-\text{sol}%
}\right \vert ^{2}dx=2A_{s}/\mu$. The velocity of soliton is affected by the
Feshbach resonance parameter $\lambda$, which reads $V_{s}=-2k_{s}(e^{\lambda
t}-\lambda e^{-\lambda t}\int_{0}^{t}e^{2\lambda \tau}\,d\tau)$.$\allowbreak$
This result show that the size of bright soliton in BECs can be tuned by
adjusting the attractive interactions even in the presence of the expulsive potential.

Another interesting problem is to discuss the interaction of two nonautonomous
bright solitons. To this purpose we present two-soliton solution of Eq.
(\ref{nls1}) as follows
\begin{equation}
\psi=\frac{G_{b}}{F_{b}}\exp \left(  \lambda t/2\right)  , \label{bso2}%
\end{equation}
where%
\begin{align*}
F_{b}  &  =f_{1}\cosh \left(  \theta_{1}+\theta_{2}\right)  +f_{2}\cosh \left(
\theta_{1}-\theta_{2}\right) \\
&  +f_{3}\cos \left(  \varphi_{1}-\varphi_{2}\right)  ,\\
G_{b}  &  =g_{1}\cosh \theta_{1}e^{i\varphi_{2}}+g_{2}\cosh \theta
_{2}e^{i\varphi_{1}}\\
&  +ig_{3}\left(  \sinh \theta_{1}e^{i\varphi_{2}}-\sinh \theta_{2}%
e^{i\varphi_{1}}\right)  ,
\end{align*}%
\begin{align*}
f_{1}  &  =\left(  k_{2}-k_{1}\right)  ^{2}+\mu^{2}\left(  A_{2}-A_{1}\right)
^{2},\\
\text{ }f_{2}  &  =\left(  k_{2}-k_{1}\right)  ^{2}+\mu^{2}\left(  A_{1}%
+A_{2}\right)  ^{2},\\
f_{3}  &  =-4\mu^{2}A_{1}A_{2},\\
g_{1}  &  =2A_{2}[\left(  k_{2}-k_{1}\right)  ^{2}+\mu^{2}\left(  A_{2}%
^{2}-A_{1}^{2}\right)  ],\\
g_{2}  &  =2A_{1}[\left(  k_{2}-k_{1}\right)  ^{2}+\mu^{2}\left(  A_{1}%
^{2}-A_{2}^{2}\right)  ],\\
g_{3}  &  =4\mu A_{1}A_{2}\left(  k_{2}-k_{1}\right)  ,
\end{align*}
with the parameters
\begin{align}
\theta_{j}  &  =\mu A_{j}\left(  xe^{\lambda t}-2k_{j}\int_{0}^{t}%
e^{2\lambda \tau}d\tau \right)  -\theta_{0j},\nonumber \\
\varphi_{j}  &  =k_{j}xe^{\lambda t}-\frac{\lambda x^{2}}{4}+(\mu^{2}A_{j}%
^{2}-k_{j}^{2})\int_{0}^{t}e^{2\lambda \tau}d\tau-\varphi_{0j}, \label{para1b}%
\end{align}
here $j=1,2$, $\theta_{0j}$ and $\varphi_{0j}$ is an arbitrary real constant,
respectively. The solution in Eq. (\ref{bso2}) describes a general interaction
between two nonautonomous solitons with the different center velocity
$V_{s,1}$ and $V_{s,2}$, respectively. From Eq. (\ref{para1b}) we get the
velocity of each soliton
\[
V_{s,j}=-2k_{s,j}(e^{\lambda t}-\lambda e^{-\lambda t}\int_{0}^{t}%
e^{2\lambda \tau}\,d\tau),j=1,2.
\]
In order to understand the nature of two nonautonomous solitons interaction,
we analyze the asymptotic behavior of two-soliton solution in Eq.
(\ref{bso2}). Asymptotically, the solution in Eq. (\ref{bso2}) can be written
as a combination of two one solutions in Eq. (\ref{bso1}). The asymptotic form
of two-soliton solution in limits $t\rightarrow-\infty$ and $t\rightarrow
\infty$ is similar to that in Eq. (\ref{bso1}).

(i) Before collision (limit $t\rightarrow-\infty$)

(a) Soliton 1 ($\theta_{1}\approx0$, $\theta_{2}\rightarrow-\infty$)%
\begin{equation}
\psi_{2-\text{sol}}\rightarrow \frac{\gamma_{1}e^{\lambda t/2}}{2\sqrt
{f_{1}f_{2}}}\frac{e^{i\left(  \varphi_{1}+\phi_{1}\right)  }}{\cosh \left(
\theta_{1}+x_{0}\right)  }, \label{asym1a}%
\end{equation}

(b) Soliton 2 ( $\theta_{2}\approx0$, $\theta_{1}\rightarrow \infty$)%
\begin{equation}
\psi_{2-\text{sol}}\rightarrow \frac{\gamma_{2}e^{\lambda t/2}}{2\sqrt
{f_{1}f_{2}}}\frac{e^{i\left(  \varphi_{2}-\phi_{2}\right)  }}{\cosh \left(
\theta_{2}-x_{0}\right)  }. \label{asym2a}%
\end{equation}

(ii) After collision (limit $t\rightarrow \infty$)

(a) Soliton 1 ($\theta_{1}\approx0$, $\theta_{2}\rightarrow \infty$)%
\begin{equation}
\psi_{2-\text{sol}}\rightarrow \frac{\gamma_{1}e^{\lambda t/2}}{2\sqrt
{f_{1}f_{2}}}\frac{e^{i\left(  \varphi_{1}-\phi_{1}\right)  }}{\cosh \left(
\theta_{1}-x_{0}\right)  }, \label{asym1b}%
\end{equation}

(b) Soliton 2 ( $\theta_{2}\approx0$, $\theta_{1}\rightarrow-\infty$)%
\begin{equation}
\psi_{2-\text{sol}}\rightarrow \frac{\gamma_{2}e^{\lambda t/2}}{2\sqrt
{f_{1}f_{2}}}\frac{e^{i\left(  \varphi_{2}+\phi_{2}\right)  }}{\cosh \left(
\theta_{2}+x_{0}\right)  }, \label{asym2b}%
\end{equation}
where
\begin{align*}
\phi_{1}  &  =\arctan \left(  g_{3}/g_{2}\right)  ,\phi_{2}=-\arg \left(
g_{3}/g_{1}\right)  ,\\
\gamma_{1}  &  =\sqrt{g_{2}^{2}+g_{3}^{2}},\gamma_{2}=\sqrt{g_{1}^{2}%
+g_{3}^{2}},x_{0}=\frac{1}{2}\ln \left(  f_{2}/f_{1}\right)  .
\end{align*}
From the above asymptotic behavior of two-soliton solution, we know that there
is no change of the amplitude for each soliton during the process of
collision. However, from Eqs. (\ref{asym1a}) to (\ref{asym2b}) we find there
is a phase exchange $2\phi_{j},j=1,2,$ and center shift $2x_{0}$ for soliton 1
and soliton 2 during collision, respectively. These results show that the
collision of two nonautonomous solitons is elastic.

\subsection{Nonautonomous bright soliton solution on nonzero background}

It is easy to find two basic solutions of Eq. (\ref{nls1}). One is $\psi=0$,
which corresponds to the vacuum particle density state. The nonautonomous
soliton solutions in Eqs. (\ref{bso1}) and (\ref{bso2}) are constructed on
this zero background to the moment. The other interesting solution of Eq.
(\ref{nls1}) is a plane wave solution%
\begin{equation}
\psi_{c}=A\exp \left(  i\varphi_{c}+\lambda t/2\right)  , \label{cw1}%
\end{equation}
with the initial amplitude $A$, wave number $k$, and $\varphi_{c}=-\lambda
x^{2}/4+kxe^{\lambda t}+\left(  2\mu^{2}A^{2}-k^{2}\right)  \int_{0}%
^{t}e^{2\lambda \tau}d\tau$. This solution can be seen as the background with
the temporal variation particle density. It will be very interesting to get
the exact soliton solution and its dynamic properties on such background.
Employing Darboux transformation \cite{Lilu,Lishuqing,Xu}, we obtain the
nonautonomous soliton solution as follows%
\begin{equation}
\psi=\left(  A+A_{s}\frac{G_{c}}{\cosh \theta+a\cos \varphi}\right)  e^{\frac
{1}{2}\lambda t+i\varphi_{c}}, \label{bso_cw1}%
\end{equation}
where%
\begin{align*}
G_{c}  &  =b_{1}\cosh \theta+\cos \varphi+i\left(  b_{2}\sinh \theta+c\sin
\varphi \right)  ,\\
\theta &  =M_{I}xe^{\lambda t}-\left[  \mu A_{s}M_{R}+\left(  k+k_{s}\right)
M_{I}\right]  \int_{0}^{t}e^{2\lambda \tau}d\tau-\theta_{0},\\
\varphi &  =M_{R}xe^{\lambda t}-\left[  \left(  k+k_{s}\right)  M_{R}-\mu
A_{s}M_{I}\right]  \int_{0}^{t}e^{2\lambda \tau}d\tau-\varphi_{0},
\end{align*}
with the parameters $b_{1}=-2\mu^{2}AA_{s}/D$, $b_{2}=-2\mu AM_{R}/D$,
$c=M_{I}/(\mu A_{s})$, $D=\mu^{2}A_{s}^{2}+M_{R}^{2}$, and $M_{R}%
+iM_{I}=[(k-k_{s}-i\mu A_{s})^{2}+4\mu^{2}A^{2}]^{1/2}$, which imply that
$M_{I}=0$ as $A_{s}=0$. Here $\theta_{0}$, $\varphi_{0}$, $A_{s}$, $k_{s}$,
$A$, and $k$ is an arbitrary real constant, respectively. From Eq.
(\ref{bso_cw1}) one can see that, as $A$ vanishes, the solution (\ref{bso_cw1}%
) reduces to one soliton solution in Eq. (\ref{bso1}). On the other hand, when
the initial amplitude of the soliton $A_{s}$ vanishes, the solution $\psi$ in
Eq. (\ref{bso_cw1}) reduces to the solution in Eq. (\ref{cw1}). Therefore, the
exact solution (\ref{bso_cw1}) describes generally the dynamics of the
nonautonomous bright soliton embedded in the temporal variation particle
density background, characterized by the envelop propagation velocity
$V_{sc}=\left(  \mu A_{s}M_{R}/M_{I}+k+k_{s}\right)  \left(  e^{t\lambda
}-\lambda e^{-t\lambda}\int_{0}^{t}e^{2\lambda \tau}\,d\tau \right)  -\lambda
e^{-t\lambda}\theta_{0}/M_{I}$.

$\allowbreak$Based on the above exact solution, we analyze in detail the
modulation instability process and the formation of spatial pattern in BECs.
As discussed in Ref. \cite{Lilu,Lishuqing}, we firstly consider a special
case, i.e., $k_{s}=k$. In this situation, there are two representative results:

(i) When $4A^{2}>A_{s}^{2}$, we have $M_{I}=0$ which implies that the soliton
velocity $V_{sc}$ becomes infinite. It is to say that the soliton in Eq.
(\ref{bso_cw1}) is completely trapped in spatial direction and undergoes the
modulation instability process \cite{Ablowitz}. Indeed, by introducing a small
quantity $\epsilon=\exp(\theta_{0})$ for $\theta_{0}<0$, and then linearizing
with respect to $\epsilon$ we have the approximation for the initial value
\begin{equation}
\psi \left(  x,0\right)  \approx \left[  \rho+\epsilon \chi \cos \left(  \mu
M_{1}x-\varphi_{0}\right)  \right]  e^{-i\lambda x^{2}/4+ikx}, \label{ini1}%
\end{equation}
where $\rho=(2A^{2}-A_{s}^{2}-iA_{s}M_{1})/(2A)$ with $\left \vert
\rho \right \vert =\left \vert A\right \vert $, $\chi=A_{s}M_{1}(M_{1}%
-iA_{s})/(2A^{2})$, and $M_{1}=\sqrt{4A^{2}-A_{s}^{2}}$. The solution of the
initial value problem of Eq. (\ref{nls1}) can be well described by the
solution (\ref{bso_cw1}) under the case of $M_{I}=0$. As a result, a small
periodic perturbation of the plane wave solution may lead to the onset of instability.

(ii) When $4A^{2}<A_{s}^{2}$, the soliton (\ref{bso_cw1}) becomes
\begin{equation}
\psi=\left(  -A+M_{2}\frac{M_{2}\cos \varphi+iA_{s}\sin \varphi}{A_{s}%
\cosh \theta-2A\cos \varphi}\right)  e^{\frac{1}{2}\lambda t+i\varphi_{c}%
},\label{bso_cw2b}%
\end{equation}
where $\theta=\mu M_{2}(xe^{\lambda t}-2k\int_{0}^{t}e^{2\lambda \tau}%
d\tau)-\theta_{0}$ and $\varphi=\mu^{2}A_{s}M_{2}\int_{0}^{t}e^{2\lambda \tau
}d\tau-\varphi_{0}$ with $M_{2}=\sqrt{A_{s}^{2}-4A^{2}}$. On the nonzero
background the nonautonomous bright soliton possesses the same properties as
that on zero background. The peak value increases and the soliton width
compresses with the increasing value of the scattering length, respectively.
The dynamic soliton evolution exhibits the periodic oscillation of the
amplitude and breather behavior due to the presence of nonvanishing background
\cite{Xu,Lishuqing}. On the other hand, under the effect of the expulsive
parabolic potential, the bright soliton can propagate in the longitudinal
direction, instead of oscillation in attractive parabolic potential. The atoms
of the bright soliton against the background can be obtained, i.e.,
$\int_{-\infty}^{+\infty}\left(  \left \vert \psi \right \vert ^{2}-\left \vert
\psi \left(  \pm \infty,t\right)  \right \vert ^{2}\right)  dx=A_{s}\left(
b_{2}^{2}+c^{2}\right)  /\left \vert M_{I}\right \vert $ $\int_{-\infty
}^{+\infty}\left(  A_{s}-2A\cosh \theta \cos \varphi \right)  /\left(  \cosh
\theta+a_{1}\cos \varphi \right)  ^{2}d\theta$, where $B=M_{R}/M_{I}$,
$\varphi=B\theta+\Delta$, and $\Delta=\mu A_{s}M_{I}\left(  B^{2}+1\right)
\int_{0}^{t}e^{2\lambda \tau}d\tau+B\theta_{0}-\varphi_{0}$. By numerically
verifying this integration we find it is a conservation quantity which
indicates that during the process of the compression of the bright soliton the
number of atoms in the bright soliton keeps invariant. Another interesting
problem is how such soliton can be created. From the expression
(\ref{bso_cw2b}), we can see that the initial wave function\ can be written as
the form $\phi=-Ae^{i\varphi_{c}}\pm iM_{2}e^{i\varphi_{c}}\operatorname{sech}%
\left(  \mu M_{2}x-\theta_{0}\right)  $ as $\varphi_{0}=\pm \pi/2,\pm
3\pi/2,\cdots$. This result shows that the solution (\ref{bso_cw2b}) can be
generated by coherently adding in quadrature a bright soliton to the background.

To better understand the properties of nonautonomous bright soliton on
nonvanishing background with the case $k_{s}\neq k$, we can decompose Eq.
(\ref{bso_cw1}) into the form%

\begin{equation}
\psi=\psi_{d}+\psi_{b}, \label{so2}%
\end{equation}
where
\[
\psi_{d}=\left(  A+A_{s}b_{1}+\frac{iA_{s}b_{2}\sinh \theta}{\cosh \theta
+a_{1}\cos \varphi}\right)  e^{\lambda t/2+i\varphi_{c}},
\]%
\[
\psi_{b}=A_{s}e^{\lambda t/2+i\varphi_{c}}\frac{\left(  1-b_{1}^{2}\right)
\cos \varphi+ic\sin \varphi}{\cosh \theta+b_{1}\cos \varphi}.
\]
As $A=0$ we have $\psi_{d}=0$ and $\psi_{b}=A_{s}\exp(-i\varphi
)\operatorname{sech}\theta_{s}$, which shows that in Eq. (\ref{so2}) the
former is zero solution, and the latter gives rise to one soliton solution for
Eq. (\ref{nls1}). With the increasing of $\left \vert A\right \vert $, a dip
starts to occur for $\psi_{d}$ describing a perturbed grey solitary wave under
the effect of nonzero background. At the same time $\psi_{b}$ exhibits the
periodic oscillation of the amplitude in propagation which can be considered
as a perturbed bright solitary wave. Therefore, the solution (\ref{bso_cw1})
describes the superposition of the perturbed dark and bright solitary waves
expressing the procedure of the pattern formation.

\section{Nonautonomous dark soliton solutions}

In this section, we consider the case of repulsive interaction between atoms,
i.e., $a_{0}>0$. Using the results in Ref. \cite{Lilu1,Hirota}, we obtain the
nonautonomous dark soliton solution of Eq. (\ref{nls1}) in the form%
\begin{equation}
\psi=\frac{\sqrt{\kappa}}{\mu}q\left(  x,t\right)  \exp \left(  i\varphi
_{d}+\lambda t/2\right)  , \label{dso}%
\end{equation}
where $\varphi_{d}=-\frac{1}{4}\lambda x^{2}+\xi_{0}xe^{\lambda t}-\left(
\xi_{0}^{2}+2\kappa \right)  \int_{0}^{t}e^{2\lambda \tau}d\tau+\zeta_{0}$ and
$q(x,t)$ is to be determined. For nonautonomous one-soliton solution, $q(x,t)$
is given by%
\begin{equation}
q\left(  x,t\right)  =\frac{1}{2}[\left(  1+Z_{1}\right)  -\left(
1-Z_{1}\right)  \tanh \frac{\eta_{1}}{2}], \label{dso1}%
\end{equation}
where
\begin{align}
\eta_{1}  &  =P_{1}xe^{\lambda t}+(\sqrt{4\kappa-P_{1}^{2}}-2\xi_{0})P_{1}%
\int_{0}^{t}e^{2\lambda \tau}d\tau+\zeta_{1},\nonumber \\
Z_{1}  &  =\frac{\sqrt{4\kappa-P_{1}^{2}}+iP_{1}}{\sqrt{4\kappa-P_{1}^{2}%
}-iP_{1}}. \label{para2a}%
\end{align}
From Eq. (\ref{para2a}) we see that the existence of nonautonomous dark
soliton implies the condition $\kappa \geq P_{1}^{2}/4$. From the solution
(\ref{dso}) with Eq. (\ref{dso1}) we clear two special cases, i.e., $Z_{1}%
=\pm1$. When $Z_{1}=1$, the nonautonomous dark solitonin Eqs. (\ref{dso})
reduces to plane wave solution $\psi=\sqrt{\kappa}/\mu e^{i\varphi_{d}+\lambda
t/2}$, which corresponds to the variation distribution density of bosons due
to the existence of the Feshbach resonance parameter $\lambda$. When
$\lambda=0,$ this case corresponds to the uniform distribution density of
bosons. On the other hand, when $Z_{1}=-1$ the solution in Eqs. (\ref{dso})
and (\ref{dso1}) becomes $\psi_{1}=-\sqrt{\kappa}/\mu e^{i\varphi_{d}+\lambda
t/2}\tanh \left(  \eta_{1}/2\right)  $, where $\eta_{1}=2\sqrt{\kappa
}xe^{\lambda t}-4\sqrt{\kappa}\xi_{0}\int_{0}^{t}e^{2\lambda \tau}d\tau
+\zeta_{1}$. This solution represents black soliton solution in BECs with
variational amplitude and the soliton velocity $V_{d}=2\xi_{0}(e^{\lambda
t}-\lambda e^{-\lambda t}\int_{0}^{t}e^{2(\lambda \tau)}\,d\tau)$.

The result in Eqs. (\ref{dso}) and (\ref{dso1}) shows that this nonautonomous
dark soliton undergo compressing effect with the increasing of the scattering
length, while the peak value has an increase. From Eqs. (\ref{dso}) and
(\ref{dso1}) we have $\left \vert q\right \vert _{\min}=\sqrt{\kappa-P_{1}%
^{2}/4}/\mu \exp \left(  \lambda t/2\right)  $ which is inverse proportion to
the initial soliton width $1/P_{1}$. The dark soliton possesses the
accelerated motion with the absolute increasing scattering length tuned by the
Feshbach resonance instead of oscillation in attractive parabolic potential
and the velocity reads $V_{d}=(2\xi_{0}-\sqrt{4\kappa-P_{1}^{2}})(e^{\lambda
t}-\lambda e^{-\lambda t}\int_{0}^{t}e^{2(\lambda \tau)}\,d\tau)$ obtained from
Eq. (\ref{para2a}). From Eq. (\ref{dso}) we obtain the particles number in the
form $\int_{-\infty}^{+\infty}\left(  \left \vert \psi \right \vert
^{2}-\left \vert \psi \left(  \pm \infty,t\right)  \right \vert ^{2}\right)
dx=-P_{1}/\mu^{2}$, which is a conservation quantity. The above analysis
implies that we can control the dip matter density of dark soliton by
adjusting appropriately the Feshbach resonance.

For nonautonomous dark two-soliton solution, the expression of $q(x,t)$ is
given by
\begin{equation}
q_{2}\left(  x,t\right)  =\frac{1+Z_{1}e^{\eta_{1}}+Z_{2}e^{\eta_{2}}%
+A_{12}Z_{1}Z_{2}e^{\eta_{1}+\eta_{2}}}{1+e^{\eta_{1}}+e^{\eta_{2}}%
+A_{12}e^{\eta_{1}+\eta_{2}}}, \label{dso2}%
\end{equation}
where%
\begin{align}
\eta_{j}  &  =P_{j}xe^{\lambda t}+(\sqrt{4\kappa-P_{j}^{2}}-2\xi_{0})P_{j}%
\int_{0}^{t}e^{2\lambda \tau}d\tau+\zeta_{1,j},\nonumber \\
Z_{j}  &  =\frac{\sqrt{4\kappa-P_{j}^{2}}+iP_{j}}{\sqrt{4\kappa-P_{j}^{2}%
}-iP_{j}},\nonumber \\
A_{12}  &  =\frac{4\kappa-P_{1}P_{2}-\sqrt{4\kappa-P_{1}^{2}}\sqrt
{4\kappa-P_{2}^{2}}}{4\kappa+P_{1}P_{2}-\sqrt{4\kappa-P_{1}^{2}}\sqrt
{4\kappa-P_{2}^{2}}}, \label{para2b}%
\end{align}
here $j=1,2$. From the solution in Eq. (\ref{dso2}), we can see its asymptotic
behavior
\begin{align*}
q_{2}\left(  x,t\right)   &  \rightarrow Z_{1}Z_{2},\text{ as }x\rightarrow
+\infty,\\
q_{2}\left(  x,t\right)   &  \rightarrow1,\text{as }x\rightarrow-\infty,
\end{align*}
which shows only a phase shift $\delta_{d,1}+\delta_{d,2}$, here $\delta
_{d,j}$ $=\arctan[2P_{j}\sqrt{4\kappa-P_{j}^{2}}/\left(  4\kappa-2P_{j}%
^{2}\right)  ],j=1,2,$ as from $x\rightarrow+\infty$ to $x\rightarrow-\infty$.
The solution in Eq. (\ref{dso}) with Eq. (\ref{dso2}) describes a general
scattering process of two dark solitary waves of BECs on the nonzero
background, characterized by the different center velocity $V_{1}$ and $V_{2}%
$, respectively. From Eq. (\ref{para2b}) we get the velocity of each soliton
as $V_{j}=(2\xi_{0}-\sqrt{4\kappa-P_{j}^{2}})(e^{\lambda t}-\lambda
e^{-\lambda t}\int_{0}^{t}e^{2(\lambda \tau)}\,d\tau)$, $j=1,2$. In order to
understand the nature of two solitons interaction, we analyze the asymptotic
behavior of the solution in Eq. (\ref{dso}) with Eq. (\ref{dso2}).
Asymptotically, the two-soliton waves in Eq. (\ref{dso2}) can be written as a
combination of two one-soliton waves in Eq. (\ref{dso1}). The asymptotic form
of two-soliton solution in limits $t\rightarrow-\infty$ and $t\rightarrow
\infty$ is similar to that of one-soliton in Eq. (\ref{dso}) with Eq.
(\ref{dso1}).

(i) Before collision (limit $t\rightarrow-\infty$)

(a) Soliton 1 ($\eta_{1}\approx0$, $\eta_{2}\rightarrow-\infty$)%
\begin{equation}
\psi \rightarrow \frac{\sqrt{\kappa}e^{i\varphi_{d}+\lambda t/2}}{2\mu}%
[1+Z_{1}-\left(  1-Z_{1}\right)  \tanh \frac{\eta_{1}}{2}], \label{asym1}%
\end{equation}

(b) Soliton 2 ( $\eta_{2}\approx0$, $\eta_{1}\rightarrow \infty$)%
\begin{equation}
\psi \rightarrow \frac{\sqrt{\kappa}Z_{1}e^{i\varphi_{d}+\lambda t/2}}{2\mu
}[1+Z_{2}-\left(  1-Z_{2}\right)  \tanh \frac{\eta_{2}+\delta_{0}}{2}].
\label{asym2}%
\end{equation}

(ii) After collision (limit $t\rightarrow \infty$)

(a) Soliton 1 ($\eta_{1}\approx0$, $\eta_{2}\rightarrow \infty$)%
\begin{equation}
\psi \rightarrow \frac{\sqrt{\kappa}Z_{2}e^{i\varphi_{d}+\lambda t/2}}{2\mu
}[1+Z_{1}-\left(  1-Z_{1}\right)  \tanh \frac{\eta_{1}+\delta_{0}}{2}],
\label{asym3}%
\end{equation}

(b) Soliton 2 ( $\eta_{2}\approx0$, $\eta_{1}\rightarrow-\infty$)%
\begin{equation}
\psi \rightarrow \frac{\sqrt{\kappa}e^{i\varphi_{d}+\lambda t/2}}{2\mu}%
[1+Z_{2}-\left(  1-Z_{2}\right)  \tanh \frac{\eta_{1}}{2}], \label{asym4}%
\end{equation}
where the center shift of dark soliton is given by $\delta_{0}=\ln A_{12}$. By
analyzing the asymptotic behavior of two-soliton solution in detail, we know
that there is no change of the amplitude for each soliton during collision,
while one should notice that the factor $\left \vert Z_{j}\right \vert =1$,
$j=1,2$, again. However, from Eq. (\ref{asym1}) to Eq. (\ref{asym4}) we find a
phase exchange $\delta_{0}$ for soliton 1 and soliton 2 during collision.
These results show that the collision of two dark solitons is elastic.

\section{Conclusions}

In summary, we present a family of nonautonomous soliton solutions of BECs
with the time-dependent interatomic interaction in an expulsive parabolic
potential. Our results show that the amplitude, width, and velocity of
nonautonomous soliton can be affected by the time-dependent atomic scattering
length, which can be tuned by the external filed from the so-called Feshbach
resonance technique. These results also provide an experimental tool for
investigating the range of validity of the one-dimensional Gross-Pitaevskii
equation with an abnormal expulsive parabolic potential. For the cases of both
attractive and repulsive interaction, the total particles number is a
conservation quantity, but the peak (dip) density of soliton can be controlled
by the Feshbach resonance parameter. The soliton solutions reported here may
be more realistic and leave scope for more physical explanation and
application in the future.$\allowbreak$

\section{\textbf{Acknowledgments}}

This work is supported by Hundred Innovation Talents Supporting Project of
Hebei Province of China, the NSF of China under Grant No. 10874038, the
Province Natural Science Foundation of Shanxi under Grant No. 2007011007, and
the Natural Science Foundation of Hebei Province under Grant No. A2008000006.

\end{document}